\documentstyle[11pt]{article}
\newcommand{\be}{\begin{eqnarray}}

\newcommand{\ee}{\end{eqnarray}}

\title{
	\begin{flushright}
	{\normalsize TPI--MINN--93--52/T \\
	NUC--MINN--93--28/T \\
        UMN--TH--1224/93 \\
	October 1993 \\}
	\end{flushright}
\bf Gluon Distribution Functions for
Very Large Nuclei at Small Transverse Momentum}
\author{
	Larry McLerran and Raju Venugopalan \\
        {\small\it Theoretical Physics Institute } \\
        {\small\it  School of Physics and Astronomy } \\
 	{\small\it  University of Minnesota} \\
        {\small\it Minneapolis, MN 55455 } \\
	 }
\date{}

\begin{document}

\maketitle

\begin{center}
{\bf Abstract}\\
\end{center}
We show that the gluon distribution function for very large nuclei may be
computed for small transverse momentum as correlation functions of an
ultraviolet finite two dimensional Euclidean field theory.  This computation is
valid to all orders in the density of partons per unit area, but to lowest
order in $\alpha_s$. The gluon distribution function is proportional to $1/x$,
and the effect of the finite density of partons is to modify the dependence on
transverse momentum for small transverse momentum.
\vfill \eject

\section{Introduction}

In a previous paper, we argued that in a limited range of transverse momentum,
for small values of Bjorken $x$, quark and gluon distributions functions
for very large nuclei might be evaluated as the solution of a weakly coupled
many body theory\cite{ven}.   This result relied heavily on the technology of
light cone quantization.\cite{soper} - \cite{mueller}
Specifically, when a parameter
\be
	\mu^2 = 1.1 A^{1/3} fm^{-2} \,,
\ee
corresponding to the density of charge
squared fluctuations per unit area, is $\mu^2 >>
\Lambda^2_{QCD}$, then the strong coupling is $\alpha_s(\mu^2) << 1$.  When the
Bjorken $x$ is $x << A^{-1/3}$, it is then valid to replace the valence quarks
by delta functions of charge along the light cone.  For $\alpha_s \mu^2 <<
q_t^2 << \mu^2$, and in this range of $x$, we computed the gluon distribution
function to lowest order in weak coupling to be
\be
	{1 \over {\pi R^2}} {{dN} \over {dxd^2q_t}} = {{\alpha_s \mu^2(N_c^2-1)}
\over \pi^2} {1 \over {xq_t^2}}\, .
\ee
The gluon distribution function per unit area was precisely the
Weizsacker--Williams distribution function for gluons scaled by the density of
charge squared fluctuations per unit area.   The physical picture corresponding
to the above formula is that the Weizsacker--Williams distribution is
generated by random fluctuation in the charge per unit area, and is
similar in spirit although different in origin than pictures used to
describe nucleus--nucleus scattering.\cite{ehtamo}-\cite{eisenberg}

The approximation of small $x$ guaranties that the central region gluons see a
source of valence quarks which are of a much smaller size than a typical
gluon wavelength as measured in a frame
co--moving with the gluon distribution at that value of Bjorken $x$.  We are
therefore in the deeply screened region.  In this kinematic region, Lipatov
enhancements of the gluon distribution function are expected to modify the
Bjorken $x$ dependence of the distribution function and take $1/x \rightarrow
1/x^{1+C\alpha_s}$ where $C$ is some constant\cite{lipatov}.  If such a small
$x$ enhancement actually occurs, then the weak coupling expansion which is
allowed will only be formal since $\alpha_s(\mu^2) \ln(1/x)$ will not be small.
It is not yet clear whether this enhancement actually takes place in the deeply
screened small $x$ region we are interested in. If it does, although the
coupling constant is weak, one will have to find a way of systematically
including the effects of this enhancement.

We will not address the small $x$ enhancement in this paper.  We shall instead
turn to another aspect of the problem--which is, computing the distribution
functions in the small $q_t$ region.  We will here work to lowest (formal?)
order in $\alpha_s$ but to all orders in $\alpha_s^2 \mu^2$.  We will show that
the correlation function which gives the gluon distribution function can be
expressed as a two dimensional Euclidean correlation function of an
ultraviolet finite field
theory.  We will find that summing to all orders in $\alpha_s^2 \mu^2$ modifies
the $q_t$ distribution function. However, to all orders in this expansion, the
gluon distribution function is proportional to $1/x$.

Before deriving these results, we shall first briefly review the results of our
earlier paper.  We recall that in the small $x$ region, the valence quarks are
Lorentz contracted to a smaller distance scale than that of the wavelength of
the co--moving gluon.  Therefore the valence quarks may be treated as being
approximately delta functions along the light cone.  We can ignore valence
quark recoil, so long as the gluons being emitted are soft and so long as the
coupling is weak. In this limit, the valence quarks may be treated as sources
of charge.

The valence quarks may also be taken to be uniformly distributed in transverse
space for sufficiently large nuclei.  In this case, the scale of variation of
the nuclear valence quark distribution along the transverse direction can be
made large compared to the typical hadronic distance scales.

The problem therefore is to compute the distribution function of gluons in the
presence of static sources of color charge localized along the light cone and
uniform in transverse space. The external current due to the source may
be represented as
\be
	J^\mu_a = \delta^{+\mu} Q_a(x^+,x_t) \delta(x^-) \, .
\ee

It was shown in our first paper that the nucleus can be broken up into regions
of transverse spatial extent such that the number of valence quarks in each
region is large.  This allows us to treat the sources of charge as classical.
We also showed that the dominant contribution to the ground state wavefunction
came from states which had large fluctuations around zero color charge, but
where the fluctuations were small compared to the total amount of charge in
each
transverse spatial region. In this limit, the fluctuations are gaussian.

The $A^{1/3}$ dependence of the gluon distribution function in this screened
region follows from the above arguments. While the color charge is screened so
that the average color charge is of order $\sqrt{N}$, where $N$ is the total
color charge in each spatial region, the coherence of the color field makes the
gluon density of order $N$.  The number of valence quarks per unit
transverse area goes as $A^{1/3}$.
The gluon density therefore goes as $A^{1/3}$--up
to corrections due to the logarithmic dependence on $A$
of the coupling constant.

The problem is therefore a simple one:  If we want to compute a ground state
correlation function, we can do it by the path integral
\be
	\int [d\rho ] [dA]~ e^{-{1 \over 2\mu^2} \rho^2} ~e^{\{iS[A]
+iA_+\rho \}}
\ee
that is, we just integrate the path integral for fixed charge around a Gaussian
fluctuating charge at each point in space.  We must therefore compute
correlation functions in a stochastic background field.

The approximation that we may treat the source as classical is only true in the
limit where the spatial regions we are looking at have a large number of quarks
in them.  In our first paper, we argued that this requires that
\be
	q_t^2 << \mu^2
\ee
We will assume that this is also the case in this paper.

In the previous paper, we evaluated the ground state gluon distribution
function in the perturbative region where both $\alpha_s$ and $\alpha_s^2
\mu^2$
were treated as small parameters.  This latter condition forced $\alpha_s^2
\mu^2 << q_t^2$.  In this paper, we will relax this condition and set up the
computation of the gluon distribution function in the soft region.

\section{The Classical Problem}

We first turn to the problem of computing the solution of the classical problem
for the gluon field in the presence of a source which is a delta function along
the light cone.  The equation of motion is
\be
	D_\mu F^{\mu \nu} = g J^\nu \,,
\ee
where $J$ is the classical light cone source.  We will
work in light cone gauge where $A_- = - A^+ = 0$.

The current $J^\mu $ only has components along the $+$ direction and is
proportional to a delta function of $x^-$.  There exists a solution of the
equations of motion for this problem, where the longitudinal component $A_+$,
which is not zero by a gauge, vanishes by the equations of motion
\be
	A_+ = -A^- = 0 \, .
\ee
The only non--zero components of the field strength therefore are the
transverse components which we require to be of the form
\be
	A_i(x) = \theta (x^-) \alpha_i (x_t) \, ,
\ee
where $\theta (x^-)=+1$ for $x^- > 0$ and
$\theta(x^-) = 0$ for $x^- < 0$.  This function is nonvanishing
only when $x^- > 0$, which since $x^- = (t-x)/\sqrt{2}$ is equivalent to
$x < t$.  This is what we expect for a classical field generated by
a source traveling close to the speed of light with $x = t$.
For $x > t$, the source has not yet arrived, and for $x < t$ the source should
produce a field.

Using the definition of $F^{i+}$ in terms of $A^i$
\be
	F^{i +} = \delta (x^-) \alpha_i \, .
\ee
If we further require that
\be
	F^{ij} = 0 \,,
\ee
(where i and j are transverse components), we see that we have a solution
of the equations of motion so long as
\be
	\nabla \cdot \alpha = g \rho(x_t) \, .
\ee
Here $\rho $ is the surface charge density associated with the current $J$.
There is no dependence on $x^-$ because we have factored out the delta
function.  The dependence on $x^+$ goes away because the extended current
conservation law,
\be
	\partial_+ Q^a + if^{abc} A_+^b Q^c = 0 \,,
\ee
is simplified by the solution of the field equations in Eq. (7) to read
\be
	\partial_+ Q^a = 0 \, .
\ee
Hence,
\be
	Q^a (x^+,x_t) = \rho^a (x_t) \, .
\ee

The condition that $F^{ij} = 0$ is precisely the condition that the field
$\alpha $ is a gauge transform of the vacuum field configuration for a two
dimensional gauge theory.  The requirement that $\nabla \cdot \alpha = g\rho $,
is a condition that fixes the gauge. For such a field configuration, the light
cone Hamiltonian
\be
	P^- = 0 \, .
\ee
This is the precise analog of what we found in our previous paper for the
Weizsacker--Williams distribution around an electron.

The field configuration which is a gauge transform of the vacuum field
configuration for a two dimensional field theory may be written as
\be
	\tau \cdot \alpha_i = -{1 \over g} U {1 \over i} \nabla_i U^\dagger\, .
\ee
We have not been able to construct explicit solutions for the above
equation for arbitrary dependence of the surface charge density
on $x_t$.

Note that the field configuration which solves this problem only has a
nontrivial dependence on $x_t$.  The dependence on $x^-$, as shown in Eq. (8)
is only through a step function,
and upon Fourier transforming gives only a factor of $1/k^+$.
As we will see in the next section, the
distribution functions associated with this field therefore, to all orders
in $\alpha_s^2 \mu^2$, have the general form
\be
	{1 \over {\pi R^2}} {{dN} \over {dx d^2k_t}} =
{{(N_c^2-1)} \over \pi^2} {1 \over x}~
{1 \over \alpha_s}H(k_t^2/\alpha_s^2 \mu^2)\, .
\ee
In our previous paper, we showed explicitly that in the weak coupling limit
we obtained the simple result
\be
	H(y) = 1/y \,,
\ee
for
\be
	1 << y << 1/\alpha_s^2 \, .
\ee
We see here that this result is true only in the weak coupling limit
where $\alpha_s^2 \mu^2 << k_t^2$.  The range is larger than that used
in our previous work.  Note also that this result is parallel to that of
finite temperature field theory where there is a non-perturbative
length scale, the magnetic screening length, which is
$\lambda \sim 1/\alpha_s T$ where $T$ is the temperature.

\section{Computing Correlation Functions}

To compute correlation functions associated with our classical solutions, we
must integrate over all $\rho $.  This is equivalent to integrating over the
transverse field with the constraint that the field must be a pure gauge, that
is,
\be
\int [d\alpha ] ~e^{-(\nabla \alpha)^2/2g^2\mu^2}~ \delta(F^{12}) \, .
\ee
We can perform the integration over the transverse field $\alpha $ as
integrations over unitary matrices in the standard way in which one goes to a
latticization of a gauge theory.  The integration measure becomes
\be
	\int [dU] \exp
         \left( -{1 \over {g^4\mu^2}}
tr \left( \nabla ( U{1 \over i}\nabla U^\dagger) \right)^2 \right)
\ee

We see that the measure for this theory is that for a two dimensional Euclidean
field theory.  The spatial variables are those of the two dimensional
transverse space of the original theory. This theory is ultraviolet
finite because of the
fact the Lagrangean is fourth order in derivatives.  The expansion parameter
for the theory is is $\alpha_s^2 \mu^2/k_t^2$.

The above analysis ignores Fadeev-Popov determinants which are generated in
transforming from the integration over $\rho$ to the integration over $\alpha $
and finally to that over $U$.  To properly define the two dimensional theory
for purposes of Monte-Carlo simulation, one may have to investigate these
determinants.  For our purpose, which is to study the scaling behavior of
expectation values, we will not take into account these determinants.   The
determinant for both integral representations in terms of $\alpha $ or in terms
of $U$ is easy to compute, and is the same for both representations.  It is the
determinant of $\vec{\nabla} \cdot \vec{D}$ where $\vec{D}$ is the covariant
derivative associated with the field $\alpha $.  This leads in two dimensions
to an ultraviolet finite modification of the above measure.

As discussed in the previous section and more explicitly in our earlier paper,
when the relevant momentum scale is $k_t^2 >> \alpha_s^2 \mu^2 $, the theory is
in the weak coupling region and may be evaluated perturbatively .  For smaller
values of $k_t^2$, we are in the strong coupling phase of the theory.  In this
phase of the theory, we expect that there should be no long range order.
Correlation functions of $x_t$ should die exponentially at large distances, or
alternatively the Fourier transform of correlation functions should go like a
constant for small momentum.  We will see that this guarantees the finiteness
of the gluon distribution functions for small momentum.

It should be easy to compute correlation functions for the above action using
lattice Monte Carlo methods.  The theory is two dimensional which should make
possible the use of large grids. The theory is Euclidean so that all quantities
of physical interest are computable as Euclidean correlation functions.  The
theory is ultraviolet finite, so that there should be no problems extrapolating
to the continuum limit.

Finally, the correlation function for the computation of the transverse
momentum dependence of the structure function must be determined. This may
be simply evaluated to be
\be
	D(k_t) = {1 \over \alpha_s} H(k_t)
\ee
where $D$ is the propagator for the two dimensional theory
\be
	(2\pi )^2 \delta^2 (k_t-q_t) \delta_{ij} D(k_t) = <\alpha_i \alpha_j>
\ee
The expression for the propagator can be easily rexpressed in terms of the link
variables $U$. As claimed, the small $k_t$ behavior of the gluon distribution
function is related to the asymptotics in coordinate space for the propagator
of the two dimensional theory.   Large distances correspond to strong coupling,
which in turn corresponds to a lack of correlations signalled by an exponential
fall off.  We therefore expect that at small $k_t$, the function $H(k_t)$ will
be finite; the gluon distribution function will be non--singular.

\section{Summary and Conclusions}

We have seen that the infrared behavior of the gluon distribution
function for small $k_t$ is determined by solving a two dimensional
field theory.  The relevant fields are gauge transforms of two dimensional
vacuum configurations.  The light cone Hamiltonian vanishes for
such configurations.  The field theory for the correlation functions
of interest is finite and involves integrating over gauge transforms
of the field with a specified weight function.

It is clear that the gluon propagator and light quark propagator in the
presence of such a background field configuration should be quite simple.  The
solutions of the small fluctuation equations are just
gauge transforms of free field solutions. To construct the propagator one must
join solutions across the discontinuity in $x^-$
generated by the source of charge.  The study of
these propagators will be the subject of later analyses.

It is also clear that the infrared behavior of these correlation functions is
computable as a lattice Monte Carlo simulation.  Realistically, it appears that
the desired accuracy might be obtained.

\section{Acknowledgments}

We acknowledge support under DOE High Energy DE-AC02-83ER40105 and DOE Nuclear
DE-FG02-87ER-40328. Larry McLerran wishes to acknowledge useful conversations
with Misha Polikarpov, Janos Polonyi and Miklos Gyulassy.

\end{document}